\documentclass[lettersize,journal]{IEEEtran}
\usepackage{amsmath,amsfonts}
\usepackage{algorithmic}
\usepackage{array}
\usepackage[caption=false,font=normalsize,labelfont=sf,textfont=sf]{subfig}
\usepackage{xcolor}
\usepackage{textcomp}
\usepackage{stfloats}
\usepackage{url}
\usepackage{verbatim}
\usepackage{graphicx}

\def\BibTeX{{\rm B\kern-.05em{\sc i\kern-.025em b}\kern-.08em
    T\kern-.1667em\lower.7ex\hbox{E}\kern-.125emX}}
\usepackage{balance}
\begin{document}
\title{Integrated Communication and Imaging: Design, Analysis, and Performances of COSMIC Waveforms}

\author{Marco Manzoni$^*$,  Francesco Linsalata$^*$, Maurizio Magarini, and Stefano Tebaldini
\\ Department of Electronics Information and Bioengineering, Politecnico di Milano, Milan, Italy \\
(correspondence: marco.manzoni@polimi.it, francesco.linsalata@polimi.it)
\thanks{This work was partially supported by the European Union under the
Italian National Recovery and Resilience Plan (NRRP) of NextGenerationEU,
partnership on ``Telecommunications of the Future'' (PE00000001
- program ``RESTART'') CUP: D43C22003080001. \\$^*$These authors equally contributed to this research} 
}

\maketitle

\begin{abstract}
This paper presents COSMIC (Connectivity-Oriented Sensing Method for Imaging and Communication), an innovative waveform design framework that integrates environmental radio imaging with robust communication capabilities. COSMIC introduces an extended orthogonality condition achieved through algebraic precoding across transmitting antennas, differentiating it from conventional time, frequency, or space multiplexing techniques. By leveraging the constrained imaging field of view relative to the signal duration, COSMIC enables the simultaneous transport of information while preserving high sensing performance. Numerical evaluations reveal that COSMIC significantly outperforms state-of-the-art methods, doubling the imaging Signal-to-Noise Ratio and substantially reducing the Integrated Side Lobe Ratio, thus demonstrating its effectiveness in combining communication and imaging functionalities.
\end{abstract}


\section{Introduction}

Wireless communication and radar have traditionally been treated as separate systems, using distinct hardware and waveforms with different goals: communication for data transmission and radar for sensing the environment. In the vision of sixth generation (6G) networks, sensing is seen as an additional feature of communication, which remains the main focus. This new approach aims to integrate radar functionality into communication systems seamlessly.
However, merging radar with communication without compromising data transmission is challenging. Several studies have explored combining radar and communication by using a single waveform to transmit data and gather sensing information simultaneously \cite{DongSensig6G2024}

Most approaches limit radar to estimating range, angle, and Doppler, key for determining a target's location and speed. However, radar also aims at imaging, which implies the formation of detailed electromagnetic images of the environment \cite{tebaldini_sensing_2022}. In imaging scenarios, the waveforms must be orthogonal. However, as with automotive radars, the orthogonality condition can be relaxed when the radar range is limited. This allows room for encoding communication symbols into the waveforms while performing imaging.

Implementing communication within radar systems requires using either pulsed or continuous-wave which are common radar signals, and embedding information without disrupting radar operation is a crucial challenge \cite{hassanien2016signaling}.
In \cite{blunt2010embedding}, communication was integrated into radar by selecting waveforms from a subset, each representing a communication symbol, decoded by the receiver. Similarly, \cite{OppermannSequence} proposed weighted pulse trains using Oppermann sequences for radar-communication systems, though phase modulation risks energy leakage outside the assigned bandwidth.
While radar systems enable long-range communication, their communication capacity is often limited by their primary focus on environmental sensing.

In communication-centric systems, radar sensing is integrated as a secondary function. Cyclic Prefix Orthogonal Frequency Division Multiplexing (CP-OFDM) is popular for integrated sensing and communication (ISAC) systems. This requires optimizing space, time, and frequency resource allocation \cite{Puccietal2022, MIMOOFDM, CodeOFDMMIMOradar}.  However, OFDM high peak-to-average power ratio can be problematic in Multiple-Input Multiple-Output (MIMO) radar applications \cite{10154042}.
Alternatives like Frequency Modulated Continuous Wave (FMCW) have been adapted for dual use, but suffer from limited spectral efficiency due to their chirp-like signals \cite{aditya2022sensing}.
Dual-domain designs combining OFDM and low power sensing signal have also been explored for mono-static range and Doppler estimation \cite{tagliaferri2023integrated}.
While these techniques provide solutions to the trade-offs between communication and sensing, none have proposed a dedicated processing chain for simultaneous radar imaging and communication. A work addressing joint communication and Synthetic Aperture Radar (SAR) imaging (JCASAR) is presented in \cite{zheng2024waveform}, which employs CP-OFDM for both target profiling and communication, although it requires complex optimization to balance performance.

This letter demonstrates that a radar imaging and communication system can be achieved by applying linear precoding and decoding techniques that exploit the degrees of freedom in imaging to enable communication.

\textbf{Paper Contributions:} The main contributions of this work are as follows:
\begin{itemize}
    \item Defining the requirments for waveform design that enables simultaneous imaging and data transmission, introducing the concept of Integrated Communication and Imaging (IC\&I).
    \item Introducing COSMIC (Connectivity-Oriented Sensing Method for Imaging and Communication), a new algebraic precoding and decoding waveform design that integrates radar imaging and communication by leveraging the degrees of freedom from limited imaging fields of view.
    \item Highlighting COSMIC's ability to produce accurate environmental imaging while ensuring the communication task is met with minimal processing. Numerical results are provided to compare COSMIC IC\&I performance with current state-of-the-art waveform solutions.
\end{itemize}

\textbf{Paper Contributions} The rest of the paper is organized as follows. Section II defines the system model, the orthogonality condition for guaranteeing IC\&I system, and the COSMIC waveform design. Section III highlights the numerical results and compare COSMIC with other waveforms design in IC\&I scenario. Section III concludes and summarizes the work.

\section{System Model and IC\&I Waveform Design}

The reference scenario involves a transmitter (Tx) equipped with a MIMO transceiver with \(N\) transmitting and \(M\) receiving antennas.  The communication Rx, equipped with a single antenna, receives the superimposed signals from the \(N\) transmitters and decodes the information in a Multi-Input-Single-Output (SIMO) fashion.
The communication Rx)is not co-located with the Tx, whereas the imaging Rx is co-located with the Tx and operates in a MIMO configuration

Each transmitting antenna sends a waveform \(s_n(t)\) over a bandwidth \(B\) for a duration \(T_\mathrm{p}\). All \(N\) waveforms are transmitted simultaneously, sharing the same frequency spectrum. 
The transmitted waveforms \(s_n(t)\) may be scattered by different points, producing echoes \(s_n(t - \tau)\) with varying delays \(\tau\). These delays are proportional to the round-trip travel time from the Tx to each scattering point and back.

\subsection{Orthogonality Condition Toward IC\&I}

The discrete-time signal is defined as \(s_n(kT_s) = \mathbf{s}_n\) for \(k = 0, 1, \ldots, K-1\), where \(K\) is the number of transmitted samples, \(T_s\) is the sampling interval, and \(n = 1, 2, \ldots, N\) indexes the antennas. Each transmitting antenna conveys communication data.

The condition for perfect separation of scattered waveforms from a \textit{point target} is given by:
\begin{equation}
\label{eq:point_orthogonal}
    \mathbf{s}_n^H \mathbf{s}_m = \sum_{k=1}^{K} s_n^*(k) s_m(k) = 0 \quad \forall \, m \neq n,
\end{equation}
where $()^H$ denotes the Hermitian transpose. This condition facilitates the design of zero-shift orthogonal waveforms, commonly employed in state-of-the-art ISAC systems \cite{MIMOOFDM, liu2018toward}.

\begin{figure}[!t]
    \centering
    \includegraphics[width=\columnwidth]{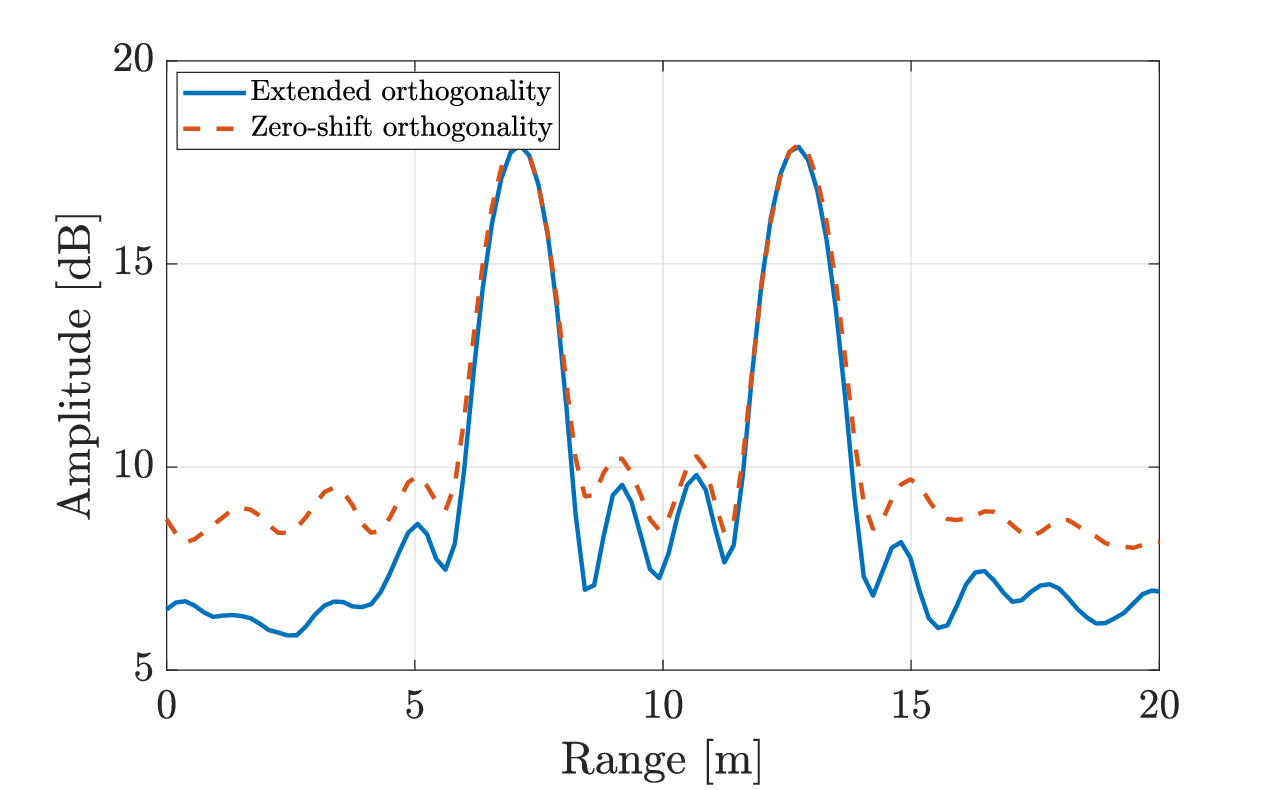}
    \caption{Average range compressed data. The scene, composed of two highly reflective targets and 100 loosely reflective ones, is illuminated by 101 waveforms simultaneously. The simulation is carried out with COSMIC waveforms and waveforms that are strictly orthogonal only for mutual shifts equal to zero.}
    \label{fig:range_compression}
\end{figure}

In a more complex scattering scenario, where an infinite number of infinitesimal scatterers are distributed at varying distances from the radar, the condition stated above is no longer \textit{sufficient}. In this case, the orthogonality requirement must be satisfied for \textit{any possible time delay} introduced by the targets in the scene.
This extended scattering orthogonality condition is expressed as:
\begin{equation}
\label{eq:extended_condition}
    \mathbf{r}_{nm} = \mathbf{S}_n \mathbf{s}_m = 0, \quad \forall \, m \neq n,
\end{equation}
where \(\mathbf{S}_n \in \mathbb{C}^{(2K-1) \times K}\) is the cross-correlation matrix, i.e. a convolution matrix constructed from the time-reversed and conjugate values of the signal \(\mathbf{s}_n\). 
Such a condition is well-known in the radar community and generally satisfied by radar waveforms to minimize MIMO noise \cite{krieger_mimo-sar_2014}. 

Figure \ref{fig:range_compression} depicts the result of a simple yet effective simulation. Two highly reflective point targets are illuminated by 108 antennas simultaneously. The signal is range-compressed (matched filtering) at the receiver and displayed. If the transmitted waveforms respect the orthogonality condition in \eqref{eq:point_orthogonal} (red line), the MIMO noise floor is much higher than the scenario in which the transmitted waveforms respect the extended orthogonality condition in \eqref{eq:extended_condition} (blue line). This means that faint targets can hardly be detected since they are immersed in noise. A figure of merit to quantify the MIMO noise can be provided by evaluating the Integrated Side Lobe Ratio (ISLR), defined as the total energy in the sidelobes divided by the energy in the main lobes.
\begin{equation}
    \mathrm{ISLR} = \frac{\int |d(r)|^2dr - \int_{\rho_r}|d(r)|^2dr} {\int_{\rho_r}|d(r)|^2dr}
\end{equation}
where $d(r)$ is the range-compressed data, $\int |d(r)|^2dr$ is the total energy over the whole range support, while $\int_{\rho_r}|d(r)|^2dr$ is the energy only within the main lobe (resolution cell). In the case of Fig. \ref{fig:range_compression}, the ISLR of the zero-shift orthogonal waveforms is -6.3 dB, while the ones providing the extended orthogonality is -7.9 dB, showing a gain of 1.6 dB.

Satisfying the requirement in \eqref{eq:extended_condition} is challenging when the intended waveform also serves communication purposes, as in an ISAC framework. Nevertheless, the key intuition behind COSMIC—and thus the main contribution of this letter—is that, in most range-limited radar imaging applications, such cross-correlation must be zero \textit{only} within the delay interval \([0, \tau_{\mathrm{s}}]\), where $\tau_{\mathrm{s}}$ is the maximum two-way travel time which is proportional to the scene size. No constraints are necessary for \(t<0\) or \(t>\tau_{\mathrm{s}}\). This consideration leads to the relaxed condition
\begin{equation}
\label{eq:orthogonality}
    \mathbf{r}_{nm} = \mathbf{\tilde{S}}_n \mathbf{s}_m = 0, \quad \forall \, m \neq n,
\end{equation}
where \(\mathbf{\tilde{S}}_n \in \mathbb{C}^{K_z \times K}\) and $K_z$ is the number of samples in the interval \([0, \tau_{\mathrm{s}}]\). Notice that $K_z \ll K$. The remaining \(K-K_z\) dimensions provide degrees of freedom for communication tasks in an IC\&I waveform design.



\subsection{Algebraic COSMIC Waveforms Design}

The \(n\)th transmitted signal is formulated as a linear combination of orthogonal bases as follows:
\begin{equation} \label{eq:signal_space}
    \mathbf{s}_n = \mathbf{C}_n \mathbf{{x}}^p_n,
\end{equation}
where \(\mathbf{{x}}^p_n \in \mathbb{C}^{K_s\times 1}\), as shown in the following, encodes the communication symbols. 

The only condition required for the matrices \(\mathbf{C}_n \in \mathbb{C}^{K \times K_s}\) to allow the communication Rx to recover the information is that they are orthogonal (i.e., $\mathbf{C}^H_n \mathbf{C}_m = \delta_{m-n}\mathbf{I}$). The number of columns of these matrices is $K_s$, and it is selected before starting the procedure for waveform generation. In a first approximation, although not mandatory, $K_s \approx K/N$. 
The matrix \(\mathbf{C}_n\) can be constructed by selecting a subset of \(K_s\) columns from a master orthogonal matrix \(\mathbf{C} \in \mathbb{C}^{K \times K}\). The multiplication by \(\mathbf{C}_n\) projects the vector \(\mathbf{x}_n^p\) into the appropriate signal space that maintains orthogonality with other signals transmitted in the system. 

Remarkably, if $\mathbf{C}$ is the Inverse DFT (IDFT) matrix,  \eqref{eq:signal_space} is comparable with an OFDM system, where the orthogonality is achieved in the frequency domain.

To ensure orthogonality as in \eqref{eq:orthogonality}, the problem can be formulated as finding $\mathbf{x}_n^p$ that minimizes
\begin{equation} \label{eq:op_prob}
\min_{\mathbf{x}_n^p} \left\| \mathbf{B}_n \mathbf{x}_n^p \right\|_2^2, 
\quad \text{subject to } \mathbf{x}_n^p \neq \mathbf{0},
\end{equation}
where $\left\| \cdot \right\|_2 $ is the norm operation and $\mathbf{B}_n = 
\begin{bmatrix} 
\tilde{\mathbf{S}}_1 \mathbf{C}_n & \tilde{\mathbf{S}}_2 \mathbf{C}_n & \dots & \tilde{\mathbf{S}}_{n-1} \mathbf{C}_n
\end{bmatrix}$.
The solution is equivalent to $\mathbf{x}_n^p$ lying in the null space of $\mathbf{B}_n$, i.e.,   $\mathbf{x}_n^p \in \mathrm{Null}(\mathbf{B}_n)$, leading to 
\begin{align}
    \mathbf{x}_n^p = \mathbf{B}_n^{\perp} \mathbf{x}_n,
\end{align}
where $\mathbf{B}_n^{\perp} \in \mathbb{C}^{K_s \times (K_s-(n-1)(K_z-1))}$ contains the orthonormal basis spanning the null space of $\mathbf{B}_n$ and  $\mathbf{x}_n \in \mathbb{C}^{(K_s-(n-1)(K_z-1)) \times 1}$ is the communication symbols vector.
Therefore, the \(n\)th COSMIC waveform can be expressed as
\begin{align} \label{eq:cosmic_waveform}
    \mathbf{s}_n = \mathbf{C}_n \mathbf{{x}}^p_n = \mathbf{C}_n \mathbf{B}_n^{\perp} \mathbf{x}_n,
\end{align}
ensuring that each new signal \(\mathbf{s}_n\) remains orthogonal to all previously transmitted signals, thereby preserving the integrity and efficiency of the communication system.

\subsection{Communication and Imaging Receivers}

At both the communication and imaging Rx, the received signal at the $m$th receiving antenna can be modeled as:
\begin{equation} \label{eq:rx_signal}
    \mathbf{y}_m = \sum_{n=1}^{N} \mathbf{H}_{m,n} \mathbf{s}_n + \mathbf{w}_m
\end{equation}
where $\mathbf{H}_{m,n}$ is the channel matrix between the $n$th transmitting and $m$th receiving antenna, and $\mathbf{w}_m$ is the noise at the $m$th Rx antenna with i.i.d. elements  $\sim\mathcal{N}_\mathbb{C}(0,\sigma^2_w)$ 

For the communication purposes, a flat-fading channel is considered due to short-range communications assumption. The $n$th channel gain \(h_n\) is mainly influenced by path loss, i.e. $h_n = -10\, \log_{10} \left( \frac{G \lambda^2}{4 \pi d^2} \right) \, \text{(dB)}$,
where \(G\) is the product of the transmit and receive antenna gains, and \(d\) is the distance between Tx and Rx. Since the Tx can be considered quasi-static over short time slots (e.g., 1 ms), the channel is assumed constant during each transmission \cite{zheng2024waveform}. 

In virtue of the orthogonality between the matrices \(\mathbf{C}_n\), the communication receiver can estimate \(\hat{\mathbf{x}}_n^p\) as
\begin{align} \label{eq:least_square}
    \hat{\mathbf{x}}_n^p &= \arg \min_{\mathbf{x}_n^p} \|\mathbf{y} - \mathbf{C}_n \mathbf{x}_n^p\|^2.
\end{align}
Solving this least-square problem leads to
\begin{align} \label{eq:zeroforcing}
    \hat{\mathbf{x}}_n^p = \left( \mathbf{C}_n^H \mathbf{C}_n \right)^{-1} \mathbf{C}_n^H \mathbf{y}.
\end{align}
Since  \(\mathbf{C}_n^H \mathbf{C}_n = \mathbf{I}\),  \eqref{eq:zeroforcing} becomes
\begin{align}
    \hat{\mathbf{x}}_n^p = \mathbf{C}_n^H \mathbf{y}= \mathbf{C}_n^H \left( \sum_{n=1}^N h_n \mathbf{s}_n + \mathbf{w} \right).
\end{align}
Then, the \(n\)th estimated transmitted waveform $\hat{\mathbf{s}}_n = \mathbf{C}_n \hat{\mathbf{x}}^p_n$ is used to construct $\hat{\mathbf{B}}_n$ and compute $\hat{\mathbf{B}}_n^{\perp}$ as in \eqref{eq:op_prob}. The estimated symbols vector is obtained as
\begin{equation}
    \hat{\mathbf{x}}_n = \left(\hat{\mathbf{B}}_n^{\perp}\right) ^H \hat{\mathbf{x}}^p_n.
\end{equation}
In case $\mathbf{C}$ coincides with the IDFT matrix, as in an MIMO radar OFDM system, it is crucial to recognize that $\mathbf{C}_n$ is identical for all $N$ antennas. As a result, recovering the vectors $\hat{\mathbf{x}}_n^p$ becomes infeasible. Indeed, the cross-talk caused by the simultaneous activity of other transmitting antennas on the same subcarrier at the same time requires code, frequency, or time division multiplexing to introduce an additional orthogonality condition \cite{MIMOOFDM, CodeOFDMMIMOradar}.

For the imaging tasks, each of the \(M\) receiving antennas, co-located with the $N$ transmitting ones, captures the delayed echoes of the transmitted signals. The term \(\mathbf{H}_{m,n}\) in \eqref{eq:rx_signal} represents the imaging channel response between the \(n\)th transmitting and \(m\)th receiving antenna, while \(\mathbf{w}_m\) denotes the noise at the \(m\)th Rx antenna. 
We consider an imaging channel where the received signal undergoes a delay \(\tau_{m,n}\), proportional to the two-way travel time between the transmitter (Tx), target, and receiver (Rx). The attenuation factor \(\alpha_{m,n}\) depends on the two-way distance \(d_{m,n}\) and the target's Radar Cross Section (RCS). 
%

At each imaging Rx, the received signals are demodulated using the transmitted carrier frequency and matched filtered with the transmitted waveform (a process commonly referred to as range compression in radar applications) 
This yields:
\begin{align}
\tilde{\mathbf{S}}_n \mathbf{y}[k] = \alpha_{nm} r_n[k - k_{m,n}] e^{-j2\pi f_0 k_{m,n} T_s},
\end{align}
where \(r_n[k]\) is the discrete autocorrelation of the \(n\)th transmitted waveform, \(k_{m,n}\) is the discrete time delay corresponding to the total round-trip propagation time between the transmitter, target, and receiver, and \(T_s\) is the sampling interval. This equality holds provided that the transmitted signals are mutually orthogonal, as defined in \eqref{eq:orthogonality}.

Each Rx antenna correlates with all transmitted signals, forming an array of \(N \times M\) virtual channels. Proper spacing of the Tx and Rx antennas ensures that the elements of the resulting monostatic virtual array are spaced by \(\lambda/4\), enabling unambiguous imaging across the entire field of view \cite{tebaldini_sensing_2022}. After pulse compression, imaging continues by back-projecting the signals onto a spatial grid, producing a high-resolution image of the environment \cite{manzoni_comparison_2022}.

\section{Simulation results}
\begin{figure*} 
    \centering
    \subfloat[\label{fig:mask}]{%
       \includegraphics[width=0.25\linewidth]{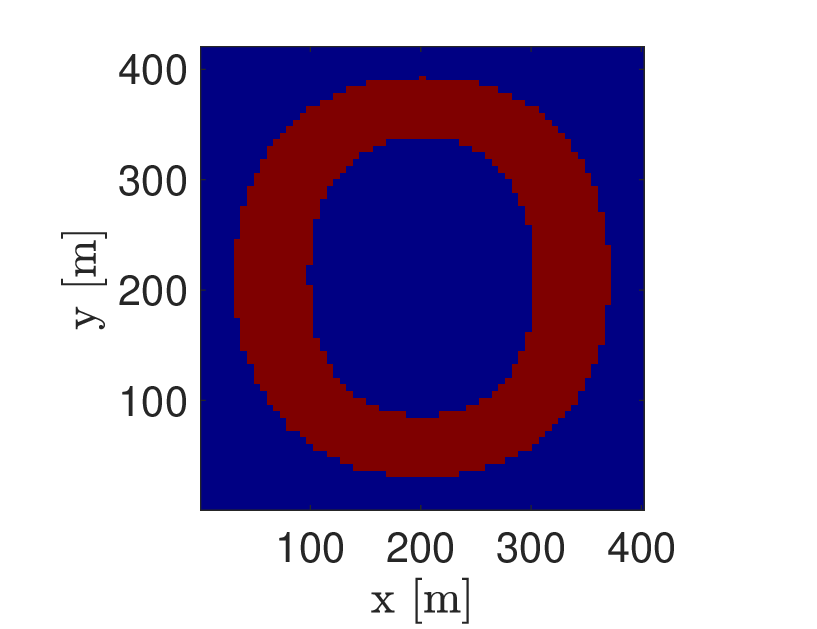}}
           \subfloat[\label{fig:image_focussed_COSMIC}]{%
       \includegraphics[width=0.25\linewidth]{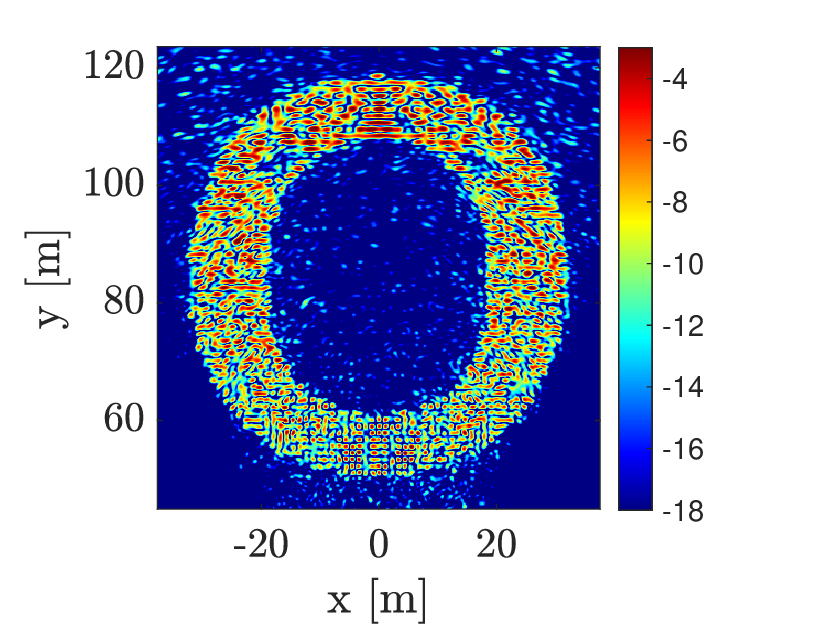}}
  \subfloat[\label{fig:image_focussed_ofdm}]{%
       \includegraphics[width=0.25\linewidth]{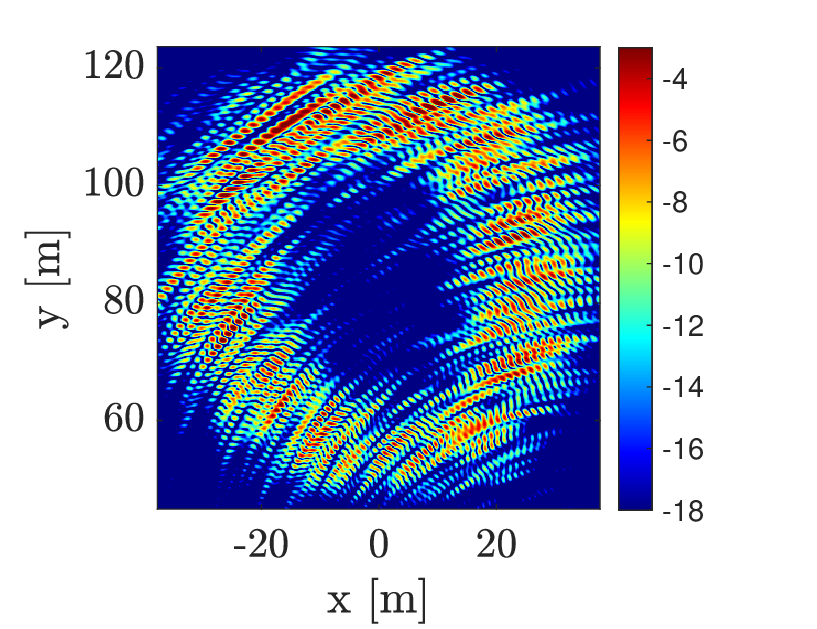}}
  \subfloat[\label{fig:image_focussed_non_orthogonal}]{%
        \includegraphics[width=0.25\linewidth]{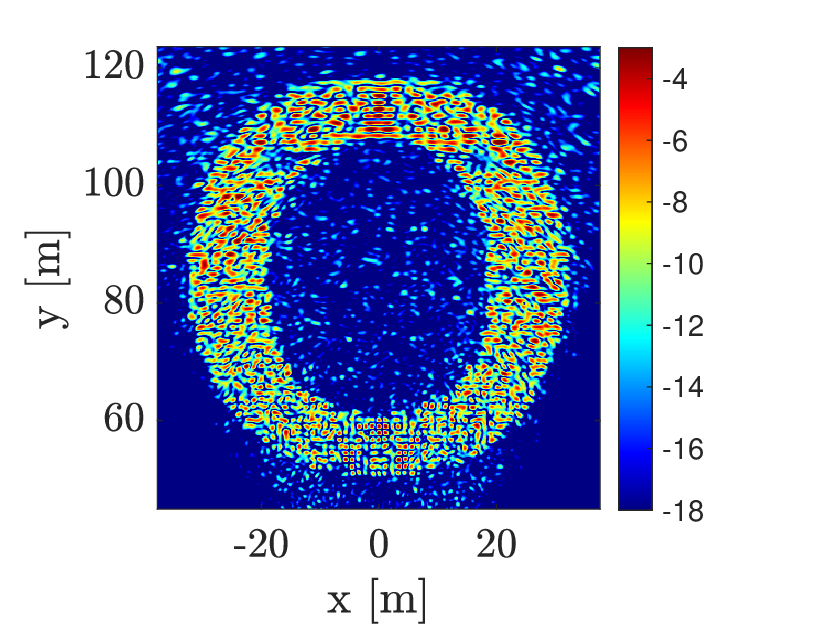}}
  \caption{Focused radar image of a) an extended scatterer by using b) COSMIC waveforms, c) MIMO radar OFDM waveform, and d) zero-shift orthogonal waveforms. The MIMO noise floor in the latter two is much higher than in the COSMIC one, leading to a higher probability of false alarms.}
  \label{fig1} 
\end{figure*}
In this section, we provide the simulation results and discuss the proposed waveform design algorithm. The simulated system exploits a bandwidth $B = 200$ MHz for a range resolution of roughly 75 cm. The antenna used for transmission and reception are $N=12$ and $M=12$, respectively, for a total of 144 virtual channels. The pulse length $T_p$ is fixed at 15 $\mu s$, while the zero-correlation zone is set at 50 m. The simulation is conducted in a noiseless environment, meaning that thermal noise is not added to the raw data. This approach is chosen to isolate and emphasize the effect of MIMO noise on the final image, which arises from the imperfect orthogonality between waveforms.

We compare the imaging performance of three different waveforms: the proposed COSMIC waveforms, MIMO radar OFDM, and a set of zero-shift orthogonal waveforms that are usually exploited in ISAC system. Each COSMIC waveform occupies the whole bandwidth, conveys information, and respects the orthogonality condition in \eqref{eq:orthogonality}. OFDM waveforms are designed to avoid that their spectrum overlap \cite{MIMOOFDM}. Finally, zero-shift waveforms span the whole bandwidth like COSMIC, but they respect only the orthogonality condition in  \eqref{eq:point_orthogonal}. Moreover, they do not carry information.

In Figure \ref{fig:mask}, the original 2D image of the environment is depicted. Figure \ref{fig:image_focussed_COSMIC} represents the image obtained using COSMIC waveforms. Unlike the one in Fig. \ref{fig:image_focussed_ofdm}, the scenario is well-focused. This behavior is expected since COSMIC waveforms span the whole bandwidth and not just a portion of it, leading to a side-lobes-free impulse response. On the contrary, each OFDM waveform spans a portion of the bandwidth and, when combined, sidelobes arise. Nevertheless, the background MIMO noise is low, as expected for orthogonal waveforms. In Figure \ref{fig:image_focussed_non_orthogonal}, the image obtained with zero-shift orthogonal waveforms is depicted. The scenario is well-focused and without sidelobes since, also in this case, each waveform covers the whole bandwidth. However, the background noise floor is much higher due to the imperfect orthogonality between the different waveforms. The increased noise floor can lead to a higher false alarm rate and may result in missed detections of faint targets within the scene.

To quantitatively compare the three waveforms, we computed the Signal-to-Noise Ratio of the image
\begin{equation}
\mathrm{SNR}_{I} = \frac{\sum_{(x, y) \in \mathcal{S}} |I(x, y)|^2}{\sum_{(x, y) \in \mathcal{N}} |I(x, y)|^2},
\end{equation}
where $I(x, y)$ represents the amplitude of the 2D image, $\mathcal{S}$ is the set of pixels where the mask in Figure \ref{fig:mask} is equal to one (signal region), and $\mathcal{N}$ is the set of surrounding pixels corresponding to the MIMO noise (noise region).
Results show that COSMIC achieves a signal-to-MIMO noise ratio of 10.3 dB, while the MIMO radar OFDM reaches only 6 dB and the zero-shift waveforms 8.8 dB.
\begin{figure} [!t]
    \centering
{
        \includegraphics[width=0.97\columnwidth]{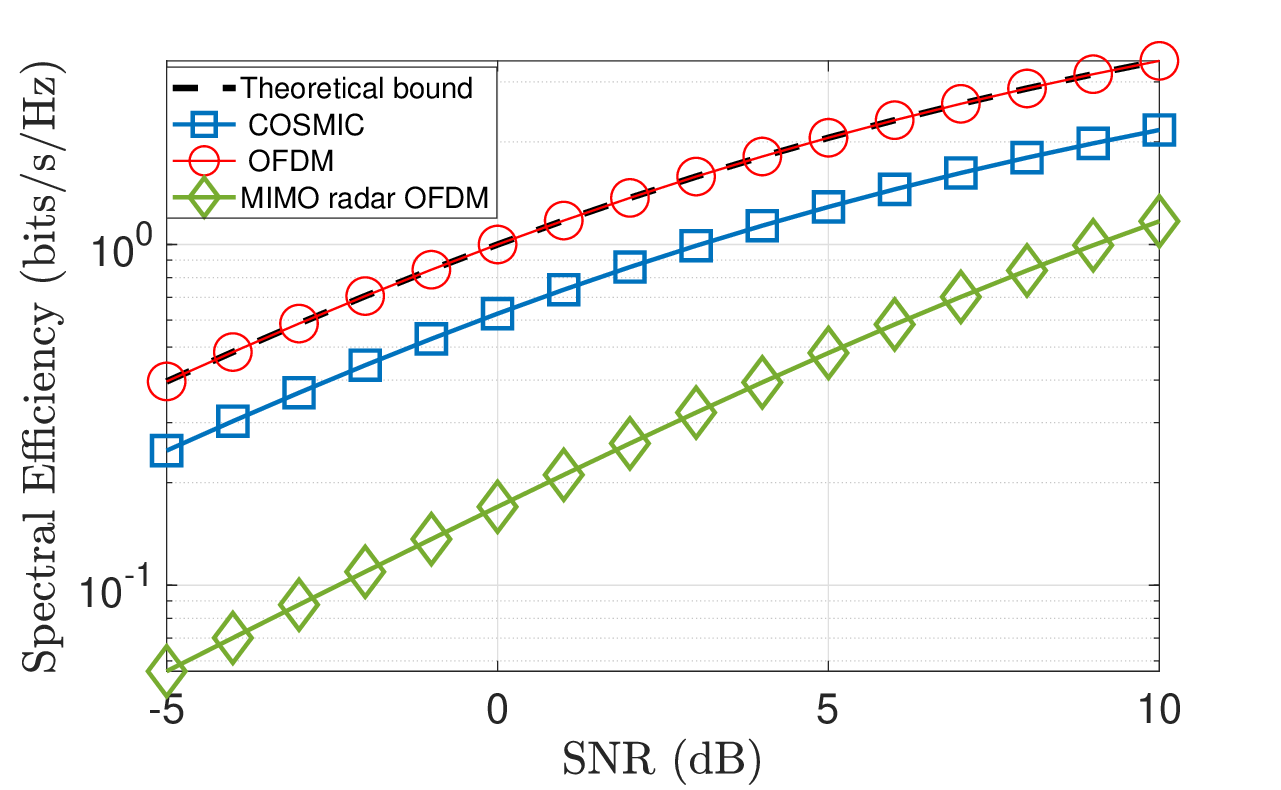} \label{fig:capacity}}      
    \caption{16-QAM Capacity versus Signal-to-Noise Ratio for COSMIC and MIMO radar OFDM waveforms.}
    \label{fig:comm_perf}
\end{figure}
The performance of the COSMIC system at the communication receiver is evaluated using the spectral efficiency (SE) in bits/s/Hz for a 16-QAM modulation scheme across different Signal-to-Noise Ratio (SNR) values, expressed in dB.
The SE is defined as
\begin{align}
\text{SE} = \beta \, \eta \sum_{n=1}^{N} \log_2 \left( 1 + \frac{|h_n|^2 P_i}{\beta\, N_0\, B} \right) \quad [\text{bits/s/Hz}]
\end{align}
where $\beta$ is the fraction of bandwidth utilized by each antenna relative to the total bandwidth $B$, $\eta$ represents the fraction of correctly received symbols, with $0 \leq \eta \leq 1$, $P_i$ is the power of the signal transmitted by the $n$-th antenna, and $N_0$ is the noise power spectral density.
The term $\frac{|h_n|^2 P_i}{N_0}$ corresponds to the SNR for each antenna, adjusted by the channel gain. 

Figure~\ref{fig:comm_perf} presents the COSMIC spectral efficiency in comparison with purely communication-oriented OFDM, the implemented MIMO radar OFDM, and the theoretical bound according to Shannon's formula. COSMIC exhibits a slight performance degradation compared to communication-oriented OFDM due to errors in estimating the null spaces at the receiver side and the reduced number of transmitted symbols over the same portion of the available bandwidth due to the constrain of the imaging. However, the COSMIC SE surpasses that of the implemented MIMO radar OFDM system, whose performance is primarily hindered by the imposed frequency orthogonality among the $N$ antennas.

\begin{figure}[!t]
    \centering
    \includegraphics[width=\columnwidth]{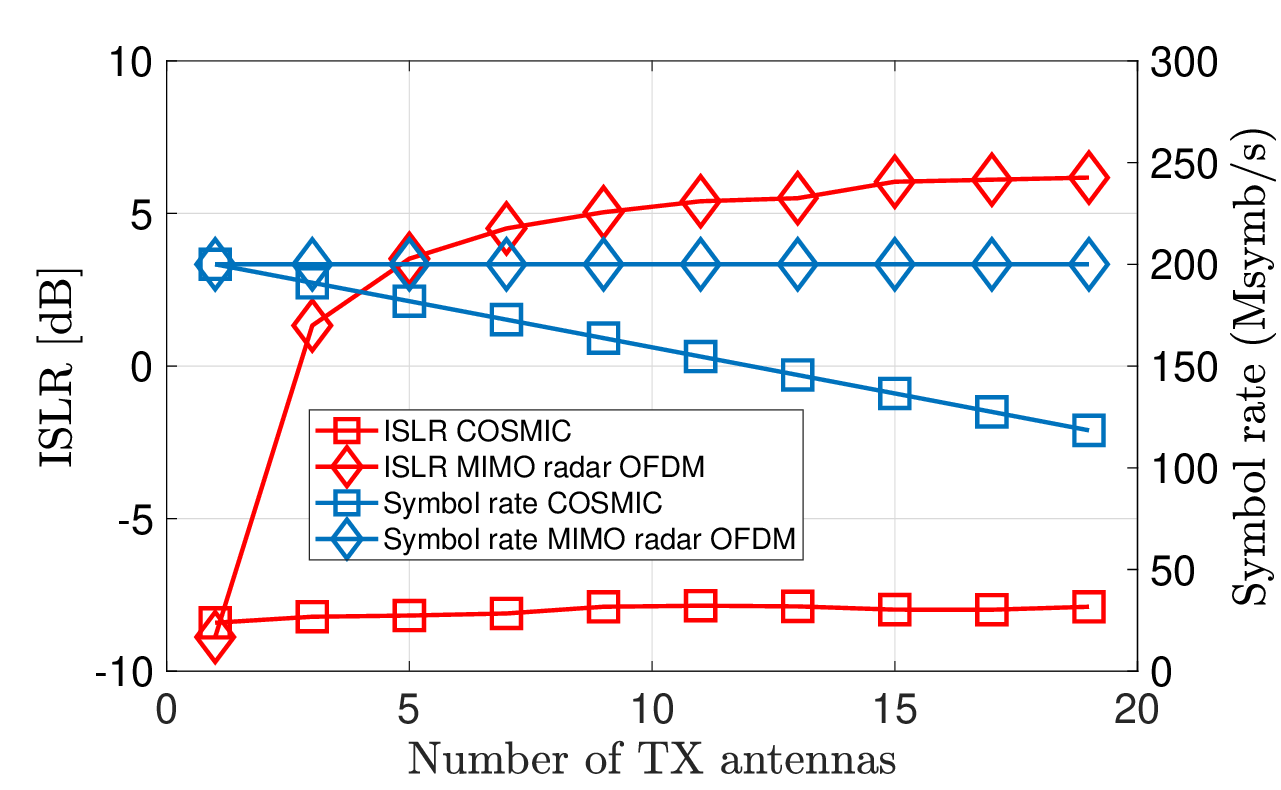}
    \caption{ISLR and achievable symbols rate versus the number of antennas $N$ for COSMIC and MIMO radar OFDM waveforms}
    \label{fig:ISLR_rate}
\end{figure}

Figure \ref{fig:ISLR_rate} illustrates an IC\&I figure of merit, comparing the ISLR and the total number of transmitted communication symbols as a function of the number of antennas \(N\) for both COSMIC and conventional MIMO radar OFDM waveforms.  
Notably, COSMIC maintains a constant ISLR, whereas the ISLR for MIMO radar OFDM increases exponentially with \(N\). This stability is achieved at the cost of linearly transmitting a lower number of communication symbols.

\section{Conclusions}

This paper introduces COSMIC (Connectivity-Oriented Sensing Method for Imaging and Communication), a novel waveform design framework for simultaneous environmental imaging and communication. Unlike traditional methods relying on time, frequency, or space multiplexing, COSMIC uses algebraic precoding across all antennas to achieve orthogonality. It leverages the limited radar imaging range to optimize waveforms for both communication and sensing without compromising performance.  
Numerical results show that COSMIC achieves a signal-to-MIMO noise ratio of 10.3 dB, compared to 6 dB for MIMO radar OFDM and 8.8 dB for zero-shift waveforms. Moreover, COSMIC improves the ISLR from \(-6.3\) dB (zero-shift) to \(-7.9\) dB, offering a gain of 1.6 dB. These findings confirm that COSMIC effectively enhances imaging performance while supporting reliable communication.

\newpage
\bibliography{references, Bibliography}
\bibliographystyle{IEEEtran}

\end{document}